# Inverse square law for the starter intensity of the discrete ordinates method in spherical geometry

*Charles H Aboughantous*\*

*Louisiana State University, Department of Physics and Astronomy, Baton Rouge, LA 70803*

***Abstract.*** *The angular derivative of the standard form of the transfer equation is shown to have a non-zero value for the radiation normal incidence. This is demonstrated by a rigorous proof. The transfer equation for the starter intensity turned out to be Euler first order equation its solution obeys the inverse square law.*

## 1. Introduction

The original discrete ordinates method in spherical geometry is structured around a set of angular parameters computed recursively from the weights and abscissas of Gauss-Legendre quadrature. The recursion relation is derived to satisfy the condition of constant asymptotic intensity [1]. The asymptotic condition is applied at the center of the sphere and recently was supplemented by the requirement that the intensity is isotropic at the center [2]. This structure produces a set of $N$ equations with $N + 1$ unknown in $S_N$ discrete ordinates. The additional unknown, labeled *starter intensity* is borrowed from the solution of the transfer equation in slab geometry for normal incidence.

The slab geometry is implemented by setting the direction cosine equal to $-1$ in the transfer equation in spherical geometry and that, as it is claimed so far, removes the angular derivative term from the equation. This starter intensity is inconsistent with the empirical flux that obeys the inverse square law. We demonstrate in this paper that the starter intensity does indeed obey the inverse square law.

## 2. The transfer equation

Consider the transfer equation in standard form for a cold gray sphere of radius $R$:

$$\mu \partial_r \psi + \frac{\eta^2}{r} \partial_\mu \psi + \sigma \psi = 0 \qquad (1)$$

where $\psi$ is used here for the specific intensity, $\partial_r$ is the tensor symbol for the derivative with respect to $r$, $\eta^2 = (1 - \mu^2)$ and $\sigma = \rho\chi$ the total extinction coefficient per unit length; $\rho$ is the density of the material assumed homogeneous and isothermal and $\chi$ the total extinction coefficient per unit mass.

Considering that Eq. (1) is the linearized form of transport equation, it is not valid at and in the close proximity of $r = 0$. Therefore, the domain of definition of Eq. (1) is $r \in [\varepsilon, R]$ and $\mu \in [-1, +1]$, where $\varepsilon$ is an arbitrarily small radius of a sphere its surface is the interior boundary for the gray medium within the limits of validity of Eq. (1).

The current practice for obtaining the starter intensity is to set $\mu = -1$ in Eq. (1). This value of $\mu$ sets $\eta = 0$ and that led to the inference that the angular derivative term vanishes. What is overseen in this cancellation process is that the quantity $(\eta^2 \partial_\mu \psi)$ must be treated as an unbreakable entity when it is evaluated at $\mu$. That is, the derivative $\partial_\mu \psi$ must be evaluated at $\mu$ simultaneously with $\eta$. Therefore, the entity $(\eta^2 \partial_\mu \psi)$ at $\mu = -1$ must be treated as a limit problem as $\mu \rightarrow -1$.

One way for obtaining the limit is to search for the same using the conservation form of the transfer equation arranged as:

$$\mu \partial_r \psi + \frac{2\mu\psi}{r} + \frac{\partial_\mu(\eta^2 \psi)}{r} + \sigma\psi = 0 \quad (2)$$

Clearly, Eq. (1) and Eq. (2) are identically the same $\forall \{r, \mu\}$ and they assume the same boundary condition. Therefore, the limit of Eq. (1) must be the same as the limit of Eq. (2) as $\mu \to \pm 1$.

### 2.1. The search for the limit

*Theorem*: $\lim_{\mu \to \pm 1}(\eta^2 \partial_\mu \psi) = \pm 2\psi$.

*Proof*. Take the limit of Eq. (1) and equate it with the limit of Eq. (2) to obtain:

$$\lim_{\mu \to \pm 1}(\eta^2 \partial_\mu \psi) = \pm 2\psi + \lim_{\mu \to \pm 1}(\partial_\mu \eta^2 \, \psi) \quad (3)$$

All we need now is to show that the limit term of the right side of Eq. (3) vanishes. It was shown that the angular derivative of the transfer equation in conservation form could be represented in terms of the angular parameters of the set $S_{2N}$ discrete ordinates [3]:

$$\partial_\mu(\eta^2 \psi) = \beta_n^n \psi_n - \beta_n^{n-1} \psi_{n-1} \quad (4)$$

where $\beta_n^n = \eta_n^2/w_n \mu_n$ and $\beta_n^{n-1} = \eta_{n-1}^2/w_n \mu_n$. The set $S_{2N}$ is structured so that $\mu_N$ is the nominal direction cosine for the normal specific intensity and $\mu_{N \to \infty} = 1$. Taking the limit of Eq. (4) we obtain:

$$\lim_{\mu \to \pm 1}(\partial_\mu \eta^2 \psi) = \lim_{N \to \infty}(\beta_N^N \psi_N - \beta_N^{N-1} \psi_{N-1}) \quad (5)$$

One property of the β-parameters is that $\beta_N^N = 0$ $\forall N$ by definition and, as $N \to \infty$, $\beta_N^{N-1} \to 0$. It follows that Eq. (5) is identically zero. Combination of this result with Eq. (3) completes the proof.

### 2.2. The starter intensity

Application of the theorem to Eq. (1) at $\mu = -1$ yields:

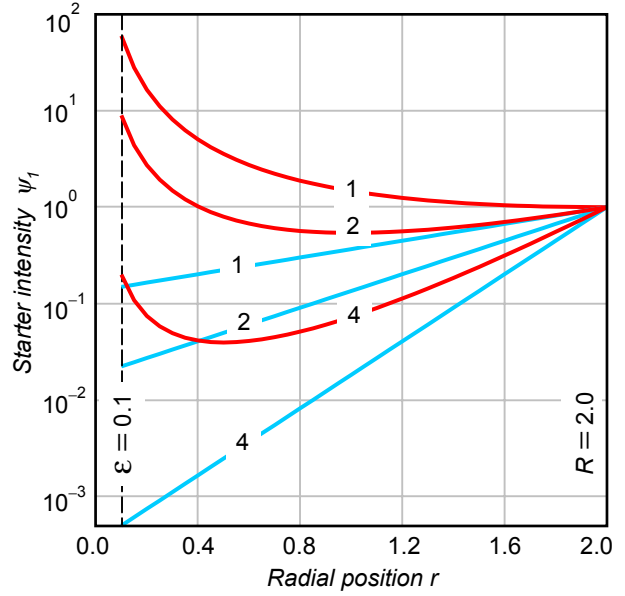

**Figure 1.** Variations of the starter intensity in spheres of different opacities σ = 1, 2, 4, shown on the curves. The graphs shown as straight lines are intensities computed in slab geometry, the curves are intensities computed with equation 7.

$$\partial_r \psi_1^- + \frac{2}{r}\psi_1^- - \sigma\psi_1^- = 0 \quad (6)$$

where the subscript 1 is for $\mu = |1|$ and $\psi_1^-$ is the normal intensity heading toward the center. We recognize that Eq. (6) is Euler first order equation, its domain is $[\varepsilon, R]$ and it assumes $\psi_R^-$ as boundary condition at the surface of the sphere. Integration of Eq. (6) on $[r, R]$ yields:

$$\psi_1^- = \frac{R^2}{r^2} e^{-\sigma(R-r)} \psi_R^- \quad (7)$$

This expression for the starter intensity is consistent with the empirical inverse square law.

It is apparent from Eq. (7) that the starter intensity in spherical geometry is $R^2/r^2$ times larger than the starter flux of slab geometry everywhere in the interior of a sphere. That is, at a point only midway between the center and the surface of the sphere, the starter intensity of slab geometry underestimates the actual radial intensity by an amount in the order of 400 percent. It follows that a solution that is conditioned by the starter intensity of slab geometry is far less accurate than it is believed.

The discrete ordinates solution of Ref. 1 is constrained to the condition of conservation of energy. Consequently, if the solution is not accurate, and if it is underestimated at some point, the condition of conservation of energy causes the solution to be overestimated at another point. That is, constraining the solution of the transfer equation to a conservation condition does not guarantee a correct solution in the interior of spheres.

Graphical representations of starter fluxes in three spheres that have outer radius $R = 2$ and inner radius $\varepsilon = 0.1$ are shown in figure 1. The opacities of the spheres are shown by numbers on the graphs. The straight lines are intensities computed in slab geometry and the curves are intensities computed with Eq. (7). The minimum of the intensity occurs at a radial position determined from Eq. (6): $r = 2/\sigma$.

## 3. Summary and conclusion

We have shown that the current practice for determining the starter intensity of the discrete ordinates method with the recursive angular parameters of Ref. 1 is not adequate. That intensity introduces large errors and the error is larger as we get closer to the center. If a proper procedure is followed, the starter intensity obeys the inverse square law consistently with the empirical method.

We note that correct starter intensity does not guarantee a correct solution with the discrete ordinates method developed around the set of recursive angular parameters of Ref. 1. Indeed, the recursion relation is conditioned by a constant asymptotic intensity. Such a condition does not prevail in the interior of a finite sphere. This is understood in light of the nature of the starter intensity that obeys the inverse square law.

These conclusions are rigorously valid in cold gray media for monochromatic radiation field. It is legitimate to argue that achromatic transfer with complete redistribution could produce results with smaller errors. The magnitude of such errors is yet to be determined.